# Models and methods of complex evaluation of complex network and hierarchically network systems

Olexandr Polishchuk

Laboratory of Modeling and Optimization of Complex Systems
Pidstryhach Institute for Applied Problems of Mechanics and Mathematics, National Academy of Sciences of Ukraine,
Lviv, Ukraine
od_polishchuk@ukr.net

**Abstract** – Current state of researches dedicated to complex network systems and different interactions between systems was analysed in this thesis. Main directions of those researches were defined and conclusions were driven that mostly those researches consist in studying structural features of systems and interactions. The flow models of complex network system were developed and main local and global flow characteristics of its elements were defined. Those characteristics are applied for determining actual structure of the system, analysing its development on all lifecycle stages and finding alternative flow directions that bypass isolated areas of network. Main structural and functional features for classification of different interactions between systems are defined. The impact of interactions on operation of interconnected complex network systems is demonstrated. Influence and betweenness parameters of components of network systems and monoflow partially overlapped multilayered systems were defined. Using those parameters, the conditions were studied under which system is resistant to negative internal and external influence. Importance of different interactions within system and between systems is determined. The scenarios are proposed of deliberate attacks on network systems, in order to define the most attractive targets of those attacks and to develop efficient ways for protecting those targets. Methods are proposed for reducing complexity of network systems and different interactions between systems which allows us to create relevant models of those systems and interactions. The notion of flow core of complex network systems and monoflow partially overlapped multilayered network systems is defined, that allows to distinguish the most important components of system structure in terms of operation, to downscale substantially system model while tracking the extent of preserving its relevance and to solve the problem of finding communities within the network. The notions of p-core and kernel of monoflow partially overlapped multilayered network are defined that allow to determine components of system structure that play the most important role in implementation of interactions between systems and to downscale respective models substantially. The main operational features of complex hierarchical network systems are defined, and methodology is developed for multiparameter and multicriteria complex evaluation of state and operation quality of those systems. The methodology proposed combines the interconnected methods of local, forecasting, interactive and aggregated analysis of system components behaviour on all hierarchy levels. Methods for interactive evaluation of system elements behaviour are developed that allow to formulate in real time indirect yet reasonable conclusions as to system components state and operation quality on the basis of results of continuous monitoring of flows within the system. Method for non-linear aggegation and efficient ways of hybridization of known aggregation procedures are developed that allow to drive much more relevant generalised conclusions as to state and operation of system components on different levels of hierarchy. It is demonstrated that application of proposed complex evaluation methodology is efficient for solving problem of defining practically possible criteria of network system element quality, novelty detection, determining critical and optimal system operation modes and selecting optimal system of given class of systems. Developed methods of complex evaluation localize priority system components requiring quality and operation improvement more accurate than existing ones and allow to timely forecast dangerous trends of development of those components and to plan expenses for eliminating potential and existing risks. Those *methods might be used while developing expert systems and decision support systems of different purposes in different subject domains.*

## ЗАГАЛЬНА ХАРАКТЕРИСТИКА РОБОТИ

**Актуальність теми.** Системи різного походження, типу та призначення є чи не найскладнішими об'єктами наукових досліджень. Це пояснюється необхідністю визначення їх складу та структури, пізнання законів функціонування та особливостей взаємодії між собою та з оточуючим середовищем великої кількості різнорідних об'єктів, які діють для досягнення спільної і часто не до кінця зрозумілої дослідчику цілі. Вивчаючи реальні складні системи, ми насправді досліджуємо побудовані на основі спостережень, експериментальних та теоретичних досліджень моделі цих систем (інформаційні, структурні, функціональні, математичні тощо). Поняття складності як для структур та систем, так і для їх моделей загалом має різний зміст. Складність структури визначається, зокрема, наявністю великої кількості елементів та зв'язків між ними. Однак, існує чимало прикладів, коли порівняно невеликі за розміром структури породжуються безумовно складними системами. Складність моделей систем визначається їх розмірністю, нелінійністю, коректністю і т. ін. Значний внесок у дослідження та вирішення проблеми складності у системних дослідженнях зробили такі вчені, як Калашников В. В., Месарович М., Садовський В. Н., Советов Б. Я., Allen L., Blanchard D. S., Chessell M., Johnson S., Grohe M., Laszlo E., Mitleton-Kelly E., Morin E., Norberg G., Northrop R. B., Odum H., Pavé A., Scott W. R., Walby S., Wilson A. G. та інші.

Одним із напрямків системних досліджень, який почав бурхливо розвиватися протягом останніх десятиліть, стало вивчення складних мережевих систем (МС). Мережеві структури є у макро- та мікросвіті. Вони є чи не найбільш розповсюдженою структурою у біологічних системах (нейронні, протеїнові, метаболічні, харчові, екологічні мережі тощо). Такі структури є звичними у людському соціумі (економічні, соціальні, фінансові, політичні, релігійні, професійні, родинні та багато інших). Предметом дослідження теорії складних мереж (ТСМ) є створення універсальних моделей мережевих структур, визначення статистичних властивостей, які характеризують їхню поведінку, та прогнозування поведінки мереж при зміні їх структурних властивостей. Однак, не менш важливими та розповсюдженими як під час дослідження фізичного світу, так і в людському суспільстві є ієрархічні структури. У соціумі ієрархія застосовується для оптимізації процесу управління та ефективної організації роботи створених людиною систем різного типу та призначення (державних, економічних, фінансових, військових тощо). Ієрархізація є методом наукового пізнання, який спрощує дослідження великих складних систем, та зручним способом структуризації знань, який впорядковує зберігання отриманих даних та полегшує пошук необхідної інформації (бібліотеки, архіви, файлова система в комп'ютерах тощо). Ієрархічний підхід є історично підтвердженим способом ефективного управління та наукового пізнання. Значний внесок у розвиток структурного напрямку системних досліджень зробили такі вчені, як Головач Ю. В., Нечипоренко В. И., Ahl V., Barabasi A.-L., Barrat A., Boccaletti S., Caldarelli G., Courgeau D., Dorogovtsev S. N., Ford L. R., Frangos J., Latora V., Mantegna R. N., Marcus C. M., Mendes J. F. F., Newman M., Prell C., Ravasz E., Simon H., Stanley H. E., Strogatz S. H., Vespignani A., Watts D. J., Wellman B. та багато інших.

Загалом як мережевий, так і ієрархічний підходи реалізують структурний напрямок системних досліджень та дозволяють частково подолати проблему складності принаймні на рівні вивчення структури системи. Однак, під час дослідження мережевих структур у них зазвичай не враховується наявність певного впорядкування чи підпорядкування складових, хоча воно безумовно існує у переважній більшості мережевих систем. З іншого боку, ієрархічні структури не враховують зв'язки між складовими одного рівня ієрархії. У той же час навіть у строго ієрархічних системах, наприклад військових, такі зв'язки є завжди. Тобто, структура реальних штучних та природних систем не вкладається у поняття «чистої» мережі або ієрархії. Складні ієрархічно-мережеві структури (СІМС), тобто структури, кожну складову певного рівня ієрархії яких можна зобразити у вигляді підмережі нижчого рівня ієрархії або підпорядкувати їй таку підмережу, більш точно та природно відображають особливості взаємодій у складних системах.

Подібність багатьох природних і штучних мережевих та ієрархічних структур допомагає у розробці універсальних методів дослідження цих структур, але не завжди процесів функціонування відповідних систем. Предметом функціонального напрямку системних досліджень є вивчення різних класів та типів систем, основних принципів та закономірностей їхньої поведінки, процесів цілеутворення, функціонування, розвитку та взаємодії із зовнішнім середовищем. При цьому структура системи розглядається спільно з функціями, які реалізуються складовими цієї структури та системою загалом, але функція має пріоритет над структурою. Однією з визначальних особливостей реально функціонуючих складних систем є рух потоків у них. В одних випадках забезпечення руху потоків є основною ціллю утворення та функціонування таких систем (транспортні, фінансові, торгівельні, соціальні мережі і т. ін.), у інших – процесом, який забезпечує їх життєдіяльність (рух крові, лімфи, нейроімпульсів у людському тілі тощо). Зупинка руху потоків може призвести до припинення існування таких систем. Теорія мережевих потоків, розвиток якої розпочався у середині 50-х років минулого століття, вирішує низку практично важливих проблем, які стосуються насамперед можливостей мережі, як структури, забезпечити рух потоків, а не особливостей процесу функціонування системи. Вивчення довільної системи потребує цілісного погляду на неї, який можна сформувати лише поєднуючи структурний та функціональний напрямки системних досліджень. Значний внесок у розвиток функціонального напрямку системних досліджень зробили Амосов Н. М., Антонов В. А., Болотін В. В., Бусленко М. П., Згуровський М. З., Іванов Ф. М., Колєсов Ю. Б., Мірошник І. В., Панкратова Н. Д., Bar-Yam Y., Bertalanffy L. V., Durauf S., Folke C., Francois C., Hillier F. S., Hinrichsen D., May R. M., Muller E., Pritchard A. J., Rosenthal H. E., Snooks G. D. та багато інших.

Будь-яка реальна система є відкритою, тобто вона взаємодіє з багатьма іншими системами. Ці взаємодії можуть мати різну природу, призначення та матеріальний носій. У ТСМ міжсистемні взаємодії описуються так званими багатошаровими структурами (БШС) різних видів. Натепер не сформовано загальноприйнятий понятійний апарат, призначений для розрізнення багатошарових мереж різних типів, не кажучи вже про їх чітку класифікацію. Найбільш вивченим різновидом міжсистемних взаємодій є так звані мультиплекси, у яких вузли однієї мережі-шару одночасно є вузлами багатьох інших мереж-шарів. Значно менш дослідженими є багатошарові структури, множини елементів яких перетинаються частково або не перетинаються взагалі. БШС можуть утворюватися у результаті взаємодії окремих систем одного типу та призначення, для яких ціль існування та способи реалізації цієї цілі є тотожними, або унаслідок взаємодії кількох різнотипних систем для реалізації певної спільної цілі або вирішення важливих для спільної життєдіяльності проблем. Значний внесок у дослідження міжсистемних взаємодій зробили Павлов Б. Л., Ровинсь- кий Р. Є., Чернишов В. Н., Barthelemy M., Bianconi G., Buldyrev S. V., Churchman C., De Domenico M., Griado R., Holme J. P., Ingarden R., Klimontovich Y. L., Lawrence W., Pflieger G., Sendina-Nadal I., Vazquez A., Willems J. C. та ін.

У випадку, коли унаслідок складності, побудувати адекватну математичну модель процесу функціонування системи не вдається, альтернативою виступають методи теорії оцінювання. Ці методи дозволяють об'єктивно аналізувати та прогнозувати стан, процес функціонування та взаємодію структурних складових МС та є потужним математичним інструментарієм у системах підтримки прийняття рішень. Методи теорії оцінювання можуть базуватися на детерміністичних, статистичних, стохастичних підходах або на їх поєднанні. Детерміністичні методи призначені для формування та прогнозування поведінки оцінок про реальний стан та якість функціонування елементів та складових вищих рівнів ієрархії СІМС. Основним недоліком цих методів є значні часові та матеріальні витрати, пов'язані з аналізом великої кількості елементів системи. У таких випадках, а також в умовах невизначеності або за неповної інформації про систему зазвичай використовують статистичні або стохастичні методи. Висновки, отримані з допомогою цих методів, привертають увагу до основних проблем функціонування системи, але у багатьох випадках не дозволяють виявляти конкретні елементи, збої в роботі яких можуть призвести до дестабілізації процесу функціонування окремих підсистем або системи загалом.

Під час розроблення методів оцінювання поведінки складних систем першочергову увагу зазвичай приділяють процедурам агрегації. Ці процедури формують узагальнені висновки про стан та процес функціонування СІМС загалом та її підсистем різного рівня ієрархії. Однак, нечіткі або однокритеріальні локальні оцінки елементів системи не дозволяють сформувати об'єктивний узагальнений висновок та побудувати точний прогноз поведінки СІМС навіть на короткострокову перспективу. Тому тільки багатокритеріальний та багатопараметричний аналіз поведінки характеристик елементів системи може бути обґрунтованою підставою для формування адекватних висновків на всіх рівнях ієрархії. На процес функціонування реальних складних систем можуть негативно впливати численні зовнішні та внутрішні чинники, що створюють ризики, які неможливо передбачити регулярними плановими дослідженнями. Тому особлива увага повинна приділятися неперервному моніторингу процесів, які перебігають у системі, та розробці методів оперативного аналізу і прогнозування таких ризиків. Результати оцінювання є об'єктивною та чи не найбільш вагомою підставою для прийняття у режимі реального часу правильних організаційних та управлінських рішень. Значний внесок у розвиток теорії оцінювання складних систем зробили Азгальдов Г. Г., Андріанов Ю. М., Воронін А. М., Дубов Ю. А., Желєзнов І. Г., Корніков В. В., Крилов О. М., Крисілов В. А., Хованов Н. В., Dombi J., Hwang C. L., Lung-Wen T., Norros L., Owen C. L., Patton M. Q., Ramsey F., Siliak D. D., Wittmuss A. та інші автори.

На підставі викладеного актуальною науково-технічною проблемою є дослідження функціональних властивостей мережевих систем та міжсистемних взаємодій різних типів, визначення способів редукції складності їх моделей та розроблення методів комплексного оцінювання стану та ефективності функціонування реальних систем різного призначення.

**Зв'язок роботи з науковими програмами, планами, темами.** Роботу виконано в рамках науково-дослідних тем Інституту прикладних проблем механіки і математики ім. Я. С. Підстригача НАН України «Розробка методів оптимізації нелінійних динамічних систем із змішаними обмеженнями на структуру керованих процесів та їх застосування в задачах робототехніки і біомеханіки» (1998 – 2002 рр., № держреєстрації 0198U002531), «Розробка методів оптимізації процесів керування та параметрів нелінійних механічних систем з активними та пасивними приводами і їх застосування в задачах робототехніки і біомеханіки» (2003 – 2006 рр., № держреєстрації 0103U000130), «Розвиток математичних моделей і методів дослідження нелінійної динаміки тонкостінних елементів конструкцій із композитних матеріалів стосовно прогнозування їх конструктивної міцності та надійності» (2007 – 2010 рр., № держреєстрації 0107U000358), «Комплекс програм для комп'ютерного моделювання ходи людини з протезованою гомілкою із врахуванням експериментальних біомеханічних даних» (2009 р., № держреєстрації 0109U003633), «Розробка методик та алгоритмів для аналізу динамічного стану тонкостінних композитних конструкцій і оптимізація режимів керування та параметрів маніпуляційних локомоційних систем стосовно задач біомеханіки» (2011 – 2013 рр., № держреєстрації 0110U004818), «Розвиток тополого-алгебраїчних, диференціально-геометричних та числових методів дослідження нелінійних динамічних систем у прикладних задачах фізики, біології та медицини» (2012 – 2016 рр., № держреєстрації 0111U009688), «Розробка математичних моделей та методів визначення динамічної поведінки композитних структур та нелінійних робототехнічних структур» (2014 – 2018 рр., № держреєстрації 0113U007684), «Розробка аналітико-числових методів моделювання та оптимізації складних динамічних систем і їх застосування у прикладних задачах» (2017 – 2021 рр., № держреєстрації 0161U008186), «Математичне моделювання оболонкових конструкцій із шаруватих композитів з дефектами та складних мережевих систем» (2019 – 2023 рр., № держреєстрації 0119U100573).

**Мета і задачі дисертаційного дослідження.** *Метою* дисертаційної роботи є дослідження властивостей складних мережевих систем та міжсистемних взаємодій різних типів і розроблення методів комплексного оцінювання стану та процесу функціонування складних мережевих та ієрархічно-мережевих систем із повністю та частково впорядкованим рухом потоків.

Відповідно до поставленої мети необхідно розв'язати такі наукові задачі:

- здійснити огляд та аналіз результатів дослідження складних мережевих систем і міжсистемних взаємодій різних типів, виділити основні підходи до оцінювання стану та якості функціонування складних систем залежно від їхнього типу і цільового призначення;
- визначити основні структурні та функціональні властивості складних мережевих систем з метою формування сукупностей локальних та глобальних характеристик, оцінювання яких надасть можливість сформувати достатньо повне та цілісне уявлення про поведінку цих систем;
- визначити основні структурні та функціональні властивості та удосконалити класифікацію міжсистемних взаємодій різних типів з метою формування сукупностей локальних та глобальних характеристик, оцінювання яких надасть можливість сформувати достатньо повне та цілісне уявлення про поведінку цих взаємодій;
- дослідити проблему редукції складності мережевих систем та міжсистемних взаємодій різних типів з метою побудови адекватних моделей цих систем та взаємодій, придатних для практичного застосування;
- визначити основні властивості складних систем з ієрархічно-мережевою структурою, як одного з важливих та широко розповсюджених різновидів міжсистемних взаємодій, та розробити методологію комплексного оцінювання їх стану та процесу функціонування з метою подальшого застосування, як математичного інструментарію, в системах підтримки прийняття рішень різного призначення;
- з метою подальшого розвитку методології комплексного оцінювання удосконалити методи інтерактивного оцінювання, як основного засобу неперервного моніторингу поведінки складних мережевих систем;
- визначити можливість застосування розробленої методології комплексного оцінювання для вирішення низки практично важливих проблем системного аналізу та теорії прийняття оптимальних рішень.

*Об'єктом дослідження* є складні мережеві та ієрархічно-мережеві системи, а також міжсистемні взаємодії різних типів.

*Предмет дослідження:* моделі та методи комплексного оцінювання поведінки складних мережевих та ієрархічно-мережевих систем.

*Методи дослідження:* для досягнення поставленої мети та розв'язання відповідних задач в роботі використовуються методи системного аналізу та теорії складних мереж; теорії оцінювання складних систем та математичної статистики, методи прогнозування та інтелектуального аналізу даних.

**Наукова новизна одержаних результатів** полягає в наступному:

**вперше**

- розроблено потокові моделі складних мережевих, ієрархічно-мережевих та монопотокових частково покритих багатошарових мережевих систем і визначено локальні та глобальні потокові характеристики їх елементів, які застосовано для встановлення реальної структури системи та аналізу її розвитку на усіх етапах життєвого циклу;
- визначено параметри впливу та посередництва складових мережевих, ієрархічно-мережевих та монопотокових частково покритих багатошарових мережевих систем, з використанням яких досліджені умови стійкості системи до негативних внутрішніх та зовнішніх впливів різних типів та визначено важливість різнорідних внутрішньо та міжсистемних взаємодій;
- визначено поняття потокової серцевини складних мережевих, ієрархічно-мережевих та монопотокових частково покритих багатошарових мережевих систем, яке дозволяє виділяти в структурі системи найважливіші з функціонального погляду її складові, суттєво зменшувати розмірність моделі системи з одночасним відстеженням міри збереження її адекватності та вирішувати проблему пошуку спільнот у мережі;
- визначено поняття $p$-серцевини та ядра монопотокової частково покритої багатошарової мережі, які дозволяють виділяти найважливіші для реалізації міжсистемних взаємодій складові її структури та суттєво зменшувати розмірність відповідних моделей;
- розроблено методику багатокритеріального та багатопараметричного комплексного оцінювання стану та якості функціонування складних ієрархічно-мережевих систем, яка поєднує взаємопов'язані методи локального, прогностичного, інтерактивного та агрегованого аналізу поведінки складових системи усіх рівнів ієрархії;
- розроблено уточнену бальну шкалу оцінювання поведінки характеристик елементів системи різних типів, яка дозволяє формувати значно точніші висновки про стан та процес функціонування елемента та принаймні частково локалізувати причини виявлених недоліків;
- розроблено методи інтерактивного оцінювання поведінки елементів системи, які на основі результатів неперервного моніторингу руху потоків мережею дозволяють в режимі реального часу формувати опосередковані, але від того не менш обґрунтовані висновки про стан складових системи та якість їх функціонування;
- розроблено метод нелінійного агрегованого оцінювання та ефективні способи гібридизації відомих агрегаційних процедур, які залежно від типу досліджуваної системи, дозволяють отримувати значно адекватніші узагальнені висновки про стан та процес функціонування її складових різних рівнів ієрархії;

**удосконалено**

- структурні та функціональні ознаки класифікації міжсистемних взаємодій різних типів та їх вплив на процес функціонування пов'язаних між собою складних мережевих систем;
- застосування методів комплексного оцінювання для вирішення проблем визначення практично досяжних критеріїв якості елементів мережевих систем, визначення критичних та вибору оптимальних режимів функціонування системи та оптимальної системи з даного класу еквівалентних систем;
- сценарії цілеспрямованих атак на мережеві, ієрархічно-мережеві та монопотокові частково покриті багатошарові мережеві системи з метою визначення найбільш привабливих цілей таких атак та розроблення відповідних засобів захисту системи;
- методи визначення реальних втрат від блокування окремих складових мережевих систем та алгоритми пошуку альтернативних шляхів руху потоків в обхід ізольованих зон мережі;

**набули подальшого розвитку**
- застосування методів теорії оцінювання в задачах пошуку «аномалій», зокрема, прихованої новизни, в інформаційних моделях складних ієрархічно-мережевих систем;
- методи U-статистик стосовно оцінювання стану та якості функціонування складних мережевих систем із частково впорядкованим рухом потоків на прикладі автотранспортної системи великого міста.

**Практичне значення одержаних результатів.** Розроблені в роботі понятійний апарат та методи дослідження складних мережевих систем та різнорідних міжсистемних взаємодій є внеском у прикладний аналіз поведінки природних та штучних систем різного типу та призначення. Запропоновані підходи до редукції складності дають можливість будувати математичні моделі систем, створення яких було неможливим із-за проблеми розмірності. Запропоновані методи ідентифікації найважливіших для процесу функціонування елементів системи дозволяють розробляти ефективніші засоби їх захисту від цілеспрямованих атак або протидіяти причинам досягнення порогу критичної завантаженості складових МС.

Запропоновані методи оцінювання значно точніше за відомі локалізують об'єкти системи, які потребують першочергового удосконалення стану або процесу функціонування та дозволяють адекватно планувати витрати, необхідні для їхньої модернізації. Розроблена в дисертаційній роботі методологія комплексного оцінювання може бути використана під час розробки експертних систем, систем підтримки прийняття рішень або бізнес-аналітики різного призначення у різноманітних предметних областях.

Результати роботи використовуються для оцінювання стану та якості функціонування біомеханічних і робототехнічних систем локомоційного типу, окремих підсистем залізничної транспортної системи різного рівня ієрархії, а також для аналізу ефективності організації навчального процесу у вищих навчальних закладах, що підтверджується відповідними актами.

**Особистий внесок здобувача.** Усі результати, які увійшли в дисертаційну роботу, одержані автором самостійно і подані у [1-61]. У працях, написаних у співавторстві, дисертантові належать: у [6, 27, 37, 39, 53] – розробка та розвиток концепції та основних складових методики комплексного детермінованого оцінювання якості функціонування складних ієрархічно-мережевих систем; у [7] – розробка методів багатопараметричного аналізу та прогнозування поведінки елементів складних систем; у [8, 20, 55] – розробка принципів висхідного та низхідного аналізу результатів агрегованого оцінювання складових системи різних рівнів ієрархії; у [9, 36, 38, 40, 44] – розробка та удосконалення методу інтерактивного оцінювання взаємодії основних елементів системи з потоками, які проходять через них; у [10, 52] – розробка методів пошуку фіктивних та прихованих елементів мережевих систем і альтернативних шляхів руху потоків в обхід ізольованих зон мережі та аналіз поведінки безмасштабних мереж на усіх етапах життєвого циклу; у [11] – введення поняття та дослідження властивостей потокової серцевини мережевої системи, а також аналіз поведінки ядер та потокових серцевин систем мультиплексного типу та їх моделей; у [12] – аналіз властивостей мережевих, ієрархічних та ієрархічно-мережевих структур, а також принципів формування інформаційних моделей і моделей оцінювання складних ієрархічно-мережевих систем; у [13, 19, 21, 41, 56] – розроблення загальних підходів до розпаралелювання послідовних методів локального, прогностичного, інтерактивного та агрегованого оцінювання складних систем, а також оптимізації методики комплексного оцінювання загалом; у [23, 24, 35] – опис послідовних алгоритмів оцінювання поведінки складних систем; у [26] – розробка уточненої бальної шкали оцінювання поведінки характеристик елементів складних систем; у [28] – аналіз можливостей застосування методів комплексного оцінювання складних систем стосовно проблематики розвитку регіонів країни; у [30, 31] – розробка підходів до аналізу та опрацювання великих обсягів даних, які описують процес функціонування складних ієрархічно-мережевих систем; у [32] – визначення та дослідження особливостей міжсистемних взаємодій різних типів; у [34] – введення та дослідження параметрів посередництва складних мережевих систем; у [45, 54] – аналіз проблеми складності у системних дослідженнях та засобів неперервного моніторингу процесу функціонування системи; у [48, 58] – приклади застосування сучасних інформаційних технологій для аналізу поведінки складних мережевих систем.

**Апробація результатів дисертації.** Основні положення і результати дисертації доповідались та обговорювались на XV Міжнародній конференції з автоматичного управління (Одеса, 2008); XI Міжнародній конференції «Моделювання та дослідження стійкості динамічних систем» (Київ, 2009); Міжнародних конференціях «Стратегія якості в промисловості і освіті» (Варна, Болгарія, 2011, 2013, 2014, 2017, 2018); II та III Міжнародних науково-технічних конференціях «Обчислювальний інтелект» (Черкаси, 2013, 2015); IV Міжнародній конференції «Knowledge – Ontology – Theory» (Новосибірськ, Росія, 2013); XX та XXI Міжнародних наукових конференціях «Сучасні проблеми прикладної математики та інформатики» (Львів, 2014, 2015); III та IV науково-технічних конференціях «Обчислювальні методи і системи перетворення інформації» (Львів, 2014, 2018); II Міжнародній конференції «Інформаційні технології та взаємодії» (Київ, 2015); X Міжнародній науковій конференції "Математичні проблеми механіки неоднорідних структур" (Львів, 2019); VI Міжнародній науково-практичній конференції «Математика в сучасному технічному університеті» (Київ, 2017); IX та X Українсько-польській науково-практичних конференціях «Електроніка та інформаційні технології» (Львів, 2017, 2018); Міжнародній науковій конференції «Сучасні проблеми механіки і математики» (Львів, 2018); Міжнародній науково-практичній конференції «Проблеми інфокомунікацій, науки та технологій» (Харків, 2018); XLVI Міжнародному науковому симпозіумі «Питання оптимізації обчислень» (Київ, 2019); Міжнародному науковому симпозіумі «Інтелектуальні рішення» (Ужгород, 2019); 9-му Міжнародному молодіжному науковому форумі «Litteris et Artibus» (Львів, 2019).

**Публікації.** За темою дисертації опубліковано 61 наукову працю [1-61], з них 24 статті [1-24] у фахових виданнях, 1 розділ у монографії [25], 7 статей [26-31] у закордонних журналах, 19 статей [33-51] у збірниках праць конференцій, 10 праць [52-61] у матеріалах і тезах конференцій. Основні результати дисертаційної роботи подано у статтях [1-32], 8 з яких входять до науково-метричних баз Scopus та Web of Sciences.

**Структура та обсяг роботи.** Дисертаційна робота складається зі вступу, семи розділів, висновків, списку використаних джерел у кількості 411 найменувань (на 38 стор.) та чотирьох додатків (на 36 стор.). Загальний обсяг дисертації – 451 сторінка. Основна частина викладена на 333 сторінках.

## ОСНОВНИЙ ЗМІСТ РОБОТИ

У **вступі** наведено загальну характеристику роботи, обґрунтовано актуальність теми дослідження та її відповідність до наукових програм. Сформульовано мету та задачі досліджень, визначено об'єкт, предмет та методи досліджень; подано наукову новизну та практичне значення одержаних результатів. Наведено відомості про використання та апробацію результатів дисертації, відзначено особистий внесок здобувача в роботах, виконаних у співавторстві.

У першому розділі **«Складні мережі та мережеві системи»** наведено огляд публікацій і проаналізовано проблему якісної та кількісної складності реальних природних і штучних систем та їх моделей різних типів. Визначено основні напрямки досліджень складних мережевих систем та коротко описано основні результати, отримані у межах структурного напрямку, який реалізується методами теорії складних мереж (СМ): наведена класифікація мережевих структур, визначено локальні та глобальні характеристики складових мережі та поняття її *k*- та *h*-серцевин; перелічені методи ідентифікації спільнот та розглянуто проблему уразливості безмасштабних мереж до цілеспрямованих атак на них і т. ін. Визначено поняття мережевої системи, як структури, що забезпечує рух потоків мережею, та описано поняття зваженої мережі, як способу наближення функціонального і структурного напрямків системних досліджень. На прикладах задач моделювання поведінки реальних мережевих систем показано, що структурні методи досліджень зазвичай не дають можливості вирішити основні функціональні задачі системного аналізу. Описано принципи побудови системних ієрархій різних типів та введено поняття ієрархічно-мережевої структури, яка більш природно описує взаємодію елементів реальних складних систем. Здійснено огляд публікацій та на прикладі багатошарових мережевих структур різних типів проаналізовано отримані на даний час результати дослідження різнорідних міжсистемних взаємодій і визначено неповноту та неоднозначність існуючого натепер понятійного апарату, який використовується для їх опису та подальшого дослідження. Проаналізовано основні підходи до вирішення проблеми підтримки прийняття рішень, які реалізуються методами теорії оцінювання. Показано, яким чином на основі багатокритеріального та багатопараметричного аналізу поведінки характеристик елементів системи та відповідних процедур прогнозування та агрегації вони дозволяють покращити об'єктивність і точність отриманих висновків на всіх рівнях ієрархії системи. Зроблено загальний опис методики комплексного оцінювання стану та якості функціонування складних ієрархічно-мережевих систем різного типу і призначення та наведено приклади її реального практичного застосування.

У другому розділі **«Потокові моделі мережевих систем»** визначаються основні функціональні характеристики елементів мережевих систем та досліджуються особливості їх поведінки.

У пункті 2.1 побудована потокова модель мережевої системи

$$\mathbf{P}(t,x) = \{\boldsymbol{\rho}(t,x), \mathbf{v}(t), \mathbf{u}(t), t \geq 0; \mathbf{V}(t), \mathbf{U}^C(t), \mathbf{U}^L(t), t \geq T\},$$

складовими якої є матриця щільності потоків, які пересуваються ребрами мережі в поточний момент часу *t*:

$$\boldsymbol{\rho}(t,x) = \{\rho_{ij}(t,x)\}_{i,j=1}^{N}, \quad x \in (n_i, n_j),$$

де $(n_i, n_j)$ – ребро, яке пов'язує вузли $n_i$ та $n_j$ мережі, $i,j = \overline{1,N}$, $t > 0$, $N$ – кількість вузлів мережі; поточна матриця суміжності об'ємів потоків, які знаходяться на ребрах мережі в момент часу *t*:

$$\mathbf{v}(t) = \{v_{ij}(t)\}_{i,j=1}^{N}, \quad v_{ij}(t) = \int\limits_{(n_i,n_j)} \rho_{ij}(t,x)dl, \quad t > 0;$$

інтегральна потокова матриця суміжності об'ємів потоків, які пройшли ребрами мережі за період $[t-T, t]$ до поточного моменту часу *t*:

$$\mathbf{V}(t) = \{V_{ij}(t)\}_{i,j=1}^{N}, \quad V_{ij}(t) = \frac{\tilde{V}_{ij}(t)}{\max\limits_{m,l=\overline{1,N}}\{\tilde{V}_{ml}(t)\}}, \quad \tilde{V}_{ij}(t) = \int\limits_{t-T}^{t} v_{ij}(\tau)d\tau, \quad t \geq T > 0;$$

матриця завантаженості ребер мережі в момент часу *t*:

$$\mathbf{u}(t) = \{u_{ij}(t)\}_{i,j=1}^{N}, \quad u_{ij}(t) = v_{ij}(t)/v_{ij}^{\max},$$

де $v_{ij}^{\max}$ – пропускна здатність ребра, яке поєднує вузли $n_i$ та $n_j$ мережі, $i,j = \overline{1,N}$, $t > 0$; інтегральні матриці завантаженості МС за період $[t-T, t]$ до моменту часу *t*:

$$\mathbf{U}^C(t) = \{U_{ij}^C(t)\}_{i,j=1}^N, \quad U_{ij}^C(t) = \max_{\tau \in [t-T,t]} \{u_{ij}(\tau)\},$$

та

$$\mathbf{U}^L(t) = \{U_{ij}^L(t)\}_{i,j=1}^N, \quad U_{ij}^L(t) = \left(\int_{t-T}^{t} u_{ij}^2(\tau)d\tau\right)^{1/2} / T, \quad t \geq T > 0.$$

Матриця щільності потоків використовується для поточного оперативного аналізу процесу функціонування МС. Матриці $\mathbf{v}(t)$ та $\mathbf{V}(t)$ дозволяють відстежувати поточні та інтегральні об'єми потоків, які проходять ребрами мережі, та аналізувати активність або пасивність окремих складових системи. Вони є особливо важливими під час прогнозування та планування роботи МС, оскільки дають можливість враховувати не лише попередні, але й поточні особливості її функціонування. Це дозволяє своєчасно реагувати на розгортання загрозливих процесів у системі. Матриці $\mathbf{u}(t)$, $\mathbf{U}^C(t)$ та $\mathbf{U}^L(t)$ дозволяють аналізувати рівень поточної та інтегральної критичної завантаженості вузлів та ребер мережі, яка може призвести до збоїв у роботі МС, а також визначати ефективність її функціонування. Загалом складові потокової моделі мережевої системи дають достатньо чітке кількісне уявлення про процес її функціонування, дозволяють аналізувати особливості та прогнозувати поведінку цього процесу, а також запобігати існуючим або потенційним загрозам.

У пункті 2.2 вводиться поняття вхідного потокового ступеня вузла $n_i$

$$W_i^{in}(t) = \sum_{j \in L_i^{in}} w_{ji}^{in}(t), \quad w_{ji}^{in}(t) = \int_{t-T}^{t} \rho_{ji}(\tau, x_i, y_{ji}(x_i))d\tau, \quad t \geq T > 0,$$

де $y = y_{ji}(x)$, $x \in [x_j, x_i]$, – рівняння кривої, яка поєднує вузли $n_j$ та $n_i$ мережі, $j \in L_i^{in}$, $L_i^{in} = \{l_i^1, ..., l_i^{d_i^{in}}\}$ – сукупність номерів суміжних з $n_i$ вузлів СМ ($d_i^{in}$ – вхідний ступінь вузла $n_i$), з яких спрямовуються потоки у вузол $n_i$, $i = \overline{1, N}$, та вихідного потокового ступеня вузла $n_i$

$$W_i^{out}(t) = \sum_{j \in L_i^{out}} w_{ij}^{out}(t), \quad w_{ij}^{out}(t) = \int_{t-T}^{t} \rho_{ij}(\tau, x_i, y_{ij}(x_i))d\tau, \quad t \geq T > 0, \quad j \in L_i^{out},$$

де $L_i^{out} = \{l_i^1, ..., l_i^{d_i^{out}}\}$ – сукупність номерів суміжних з $n_i$ вузлів СМ ($d_i^{out}$ – вихідний ступінь вузла $n_i$), у які спрямовуються потоки з вузла $n_i$, $i = \overline{1, N}$.

Потокові ступені є локальними динамічними характеристиками вузлів МС та вносять функціональну впорядкованість пріоритетності вузлів з однаковими структурними ступенями. А саме, пара $(W_i^{in}(t), W_i^{out}(t))$ визначає потокову завантаженість (функціональну важливість) вузла $n_i$, $i = \overline{1, N}$, у сукупності усіх вузлів мережі.

У пункті 2.3 визначаються кількісні показники складності моделей МС та аналізуються їх позитивні (збільшення кількості найкоротших шляхів сприяє більш ефективній організації руху потоків мережею, зменшенню середньої завантаженості кожного ребра, розширенню множини альтернативних шляхів руху потоків і т. ін.) та негативні (пришвидшення поширення інфекційних хвороб та комп'ютерних вірусів тощо) впливи на процес функціонування системи.

У пунктах 2.4 та 2.5 з використанням потокових характеристик елементів МС пропонуються методи пошуку фіктивних та прихованих елементів системи, ідентифікація яких сприяє встановленню реальної структури МС, зменшенню розмірності та посиленню адекватності моделі системи.

У пункті 2.6 визначаються параметри вхідного та вихідного впливу елементів мережевих систем та досліджуються особливості їх поведінки залежно від типу системи та процесів, які у ній перебігають. До параметрів вхідного впливу МС на вузол $n_i$ відносяться: сила вхідного впливу $\xi_i^{in}(t)$, яка дорівнює питомій вазі потоків в системі, які приймаються у вузлі $n_i$; область вхідного впливу $G_i^{in}(t)$, до складу якої входять усі вузли-генератори, з яких потоки спрямовуються у вузол $n_i$; потужність вхідного впливу $p_i^{in}(t)$, яка дорівнює кількості елементів множини $G_i^{in}(t)$, та діаметр вхідного впливу $\Delta_i^{in}(t)$, який дорівнює діаметру області $G_i^{in}(t)$, $i=\overline{1,N}$. Аналогічно визначаються параметри $\xi_i^{out}(t)$, $R_i^{out}(t)$, $p_i^{out}(t)$ та $\Delta_i^{out}(t)$ вихідного впливу вузла $n_i$, $i=\overline{1,N}$, на мережеву систему. Вхідну та вихідну потужність вузла можна вважати глобальними кількісними аналогами його вхідного та вихідного ступеня як у мережі, так і в системі. Проаналізовано, яким чином вхідні та вихідні параметри впливу дозволяють відстежувати динаміку зміни важливості вузла $n_i$, $i=\overline{1,N}$, у мережевій системі.

У пункті 2.7 за допомогою вхідних та вихідних параметрів впливу визначається важливість транзитних вузлів у МС, а також сила переважного впливу $\psi_i(t)$ вузла $n_i$, яка для нетранзитних вузлів системи обчислюється за співвідношенням

$$\psi_i(t) = \frac{\xi_i^{in}(t) - \xi_i^{out}(t)}{\xi_i^{in}(t) + \xi_i^{out}(t)}, \; \psi_i \in [-1, 1],$$

та дозволяє визначити, переважаючим є вплив вузла $n_i$ на МС чи навпаки, $i=\overline{1,N}$.

У пункті 2.8 визначаються параметри посередництва елементів мережевих систем, до складу яких для ребра $(n_i, n_j)$ відносяться: міра посередництва $\Phi_{ij}(t)$, яка дорівнює питомій вазі потоків в системі, які проходять ребром $(n_i, n_j)$; область посередництва $L_{ij}(t)$, до складу якої входять усі вузли, через які проходять потоки, що проходять ребром $(n_i, n_j)$; потужність посередництва $\eta_{ij}(t)$, яка дорівнює кількості елементів множини $L_{ij}(t)$ та діаметр посередництва $\Delta_{ij}(t)$, який дорівнює діаметру області $L_{ij}(t)$, $i,j=\overline{1,N}$.

До параметрів посередництва вузла $n_i$, $i,j=\overline{1,N}$, мережевої системи відносяться: міра посередництва $\Phi_i(t)$, яка дорівнює питомій вазі об'ємів потоків в системі, які проходять через вузол $n_i$; область посередництва $M_i(t)$, до складу якої входять усі вузли, через які проходять потоки, що проходять вузлом $n_i$; потужність посередництва $\eta_i(t)$, яка дорівнює кількості елементів множини $M_i(t)$ та діаметр посередництва $\Delta_i(t)$, який дорівнює діаметру області $M_i(t)$, $i=\overline{1,N}$. Параметри міри, області, потужності та діаметра посередництва вузла або ребра МС є глобальними характеристиками його важливості у процесі функціонування системи. Вони визначають, яким чином блокування цього елемента вплине на роботу області його посередництва, величину цієї області і, унаслідок цього, – всієї системи. Якщо параметри впливу визначають, якою мірою окремий вузол-генератор потоків впливає на роботу системи та система впливає на вузол-приймач, то параметри посередництва враховують взаємодію між усіма прямо та опосередковано пов'язаними, у тому числі транзитними, елементами системи. Параметри посередництва вузлів та ребер МС дозволяють визначити втрати, які очікують систему у результаті блокування цих елементів та необхідність перерозподілу руху потоків мережею.

У пункті 2.9 вводяться поняття параметрів впливу та посередництва підсистем МС, які є глобальними характеристиками їх важливості у процесі функціонування системи та визначають, яким чином блокування окремої підсистеми вплине на роботу області її посередництва, величину цієї області і, унаслідок цього, – всієї системи. Окрім того, показано, що аналіз значень параметрів посередництва підсистеми дозволяє ідентифікувати спільноти у межах МС.

У пункті 2.10 визначаються інтегральні показники процесу функціонування мережевих систем, зокрема, сумарні об'єми потоків, які проходять мережею за певний проміжок часу, та ефективність функціонування системи протягом цього проміжку порівняно з її потенційними можливостями. Визначаються параметри впливу та посередництва системи в процесі її взаємодії з системами того ж типу та призначення, які дозволяють вивчати структурні та функціональні особливості міжсистемних взаємодій.

У третьому розділі **«Потокові серцевини та деякі задачі моделювання поведінки мережевих систем»** вводиться поняття потокової серцевини МС та показується, що порівняно з $k$- та $h$-серцевинами СМ вона значно адекватніше відображає структури мережі, важливі для вивчення процесу функціонування системи.

У пункті 3.1 потокова $\lambda$-серцевина мережевої системи визначається, як найбільша підмережа вихідної мережі, для якої усі елементи інтегральної потокової матриці суміжності $\mathbf{V}(t)$ мають значення

$$V_{ij}(t) \geq \lambda, \; i, j = \overline{1, N}, \; t \geq T, \lambda \in [0, 1].$$

На низці прикладів реальних мережевих систем показується, що структури $k$- та $\lambda$-серцевин суттєво відрізняються, причому потокова $\lambda$-серцевина МС дає значно важливішу інформацію для системних досліджень, ніж $k$-серцевина її структури. Складність моделей МС, пов'язану з великою розмірністю та щільністю їх мережевої структури, можна зменшити, досліджуючи не всю систему, а лише її $\lambda$-серцевину. Так, зображена на рис. 1а залізнична транспортна мережа західного регіону України з виключеними вузлами-транзитерами містить 29 вузлів та 62 ребра (загалом до її складу входить 354 вузли), а її 4-серцевина (рис. 1б) – 12 вузлів та 35 зв'язків (зменшення у 3,42 та 1,77 разів). У той же час 0,7-серцевина цієї МС (рис. 1в, на якому товщина ліній є пропорційною до об'ємів потоків, що проходять ребрами мережі), яка забезпечує більше 80% усіх перевезень, нараховує 4 вузли та 12 зв'язків (зменшення у 7,25 та 5,17 разів відповідно). Введення поняття $\lambda$-серцевини робить практично можливим розв'язання задач керованості та спостережуваності, які для всієї залізничної транспортної системи (ЗТС) України розв'язати не вдається через надмірні обчислювальні витрати.

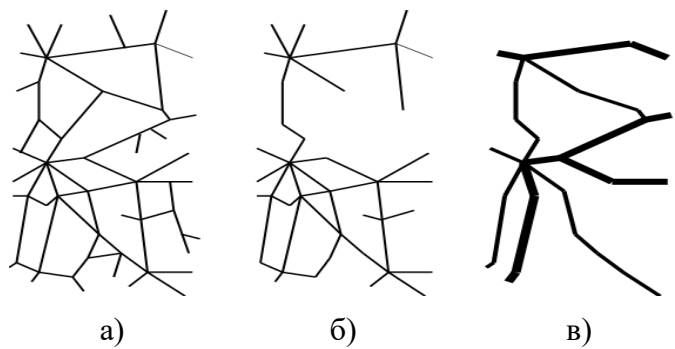

а)       б)       в)

Рис. 1. Фрагмент залізничної транспортної мережі західного регіону України (а), її структурної 4-серцевини (б) та потокової 0,7-серцевин (в)

У пункті 3.2 визначаються показники структурної та функціональної складності моделей потокових серцевин порівняно з моделлю МС загалом. Функціональна питома вага $\sigma_\lambda(t)$ потокової $\lambda$-серцевини у МС визначається з її інтегральної потокової матриці суміжності

$$\mathbf{V}^\lambda(t) = \{V_{ij}^\lambda(t)\}_{i,j=1}^N, \quad V_{ij}^\lambda(t) = \begin{cases} V_{ij}(t), \text{ якщо } V_{ij}(t) \geq \lambda \\ 0, \text{ якщо } V_{ij}(t) < \lambda \end{cases}, \quad \lambda \in [0,1], t \geq T,$$

за співвідношенням

$$\sigma_\lambda(t) = \sum_{i,j=1}^N V_{ij}^\lambda \bigg/ \sum_{i,j=1}^N V_{ij},$$

тобто дорівнює відношенню об'ємів потоків, які проходять $\lambda$-серцевиною, до об'ємів потоків, які проходять мережею загалом за період $[t-T, t]$, $t \geq T$. Оскільки основною ціллю існування або необхідною умовою життєдіяльності більшості мережевих систем є забезпечення руху певного типу потоків, то параметр $\sigma_\lambda(t)$ кількісно визначає наскільки $\lambda$-серцевина забезпечує реалізацію цієї цілі або виконання умови. Якщо замість моделі всієї системи ми досліджуємо модель її $\lambda$-серцевини, то значення параметра $\sigma_\lambda(t)$ можна інтерпретувати, як динамічну міру адекватності цієї моделі. При цьому розмірність матриці $\mathbf{V}^\lambda(t)$ з видаленими нульовими стрічками та стовпцями дорівнює розмірнісній, а кількість її ненульових елементів – коннекційній структурній складності моделі $\lambda$-серцевини.

У пункті 3.3 розглядається спосіб доповнення моделей $\lambda$-серцевин, який полягає в інкапсуляції окремих підмереж, які входять до складу решти частини вихідної МС, та їх заміщенні у структурі СМ так званими мета-вузлами. Визначаються основні ознаки підмереж вихідної МС, які можуть бути інкапсульованими, та показується, що включення відповідних мета-вузлів та їх зв'язків у структуру $\lambda$-серцевини дозволяє суттєво посилити адекватність її моделі.

У пункті 3.4 $\lambda$-серцевини розглядаються, як підсистеми вихідної МС, значення яких є значно важливішим для системи, ніж значення підсистем, сформованих за принципами ієрархії різних видів. Тому ураження потокових серцевин призводить до значно серйозніших проблем для процесу функціонування системи. Вводяться параметри впливу та посередництва $\lambda$-серцевини, як глобальні характеристики її важливості у процесі функціонування МС, та визначається, яким чином блокування цієї підсистеми вплине на роботу області її посередництва, величину цієї області і, унаслідок цього, – всієї системи.

У пункті 3.5 пропонується метод пошуку спільнот у мережі, який ґрунтується на тому, що одним із найбільш об'єктивних показників сили зв'язку між двома вузлами мережі є об'єми потоків, які проходять через поєднуюче їх ребро протягом певного періоду часу $[t-T, t]$. Це означає, що якщо під час побудови $\lambda$-серцевини вихідної МС (рис. 2а – вихідна СМ, 2б – вихідна МС з відображеними об'ємами руху потоків мережею) з послідовним зростанням $\lambda$ при певному значенні $\lambda = \lambda_1$ потокова $\lambda_1$-серцевина поділяється на незв'язні складові (рис. 2в), то виділено найбільші за своїм розміром спільноти в системі. Що важливо, структура та склад вузлів і зв'язків цих спільнот очевидним чином визначається з матриці $\mathbf{V}(t)$, $t \geq T$. Якщо з подальшим зростанням значення $\lambda$ при певному $\lambda = \lambda_2$ виділені на попередньому кроці спільноти знову поділяються на незв'язні складові, отримуємо підспільноти цих спільнот (рис. 2г) і т. д.

У пункті 3.6 досліджуються проблеми стійкості МС до цілеспрямованих атак та критичного завантаження складових системи. На основі значень параметрів впливу та посередництва елементів МС сформовані значно дієвіші за існуючі сценарії цілеспрямованих атак на систему. Ці сценарії враховують необхідність заміщення заблокованих вузлів та ребер МС і пошуку альтернативних шляхів руху потоків, які проходили через заблоковані елементи, тобто відповідний перерозподіл руху потоків мережею. Вони також дозволяють оцінити, на яку частину МС розповсюдяться наслідки збоїв відповідного елемента системи і до яких втрат це призведе у сенсі недопостачання певних об'ємів потоків.

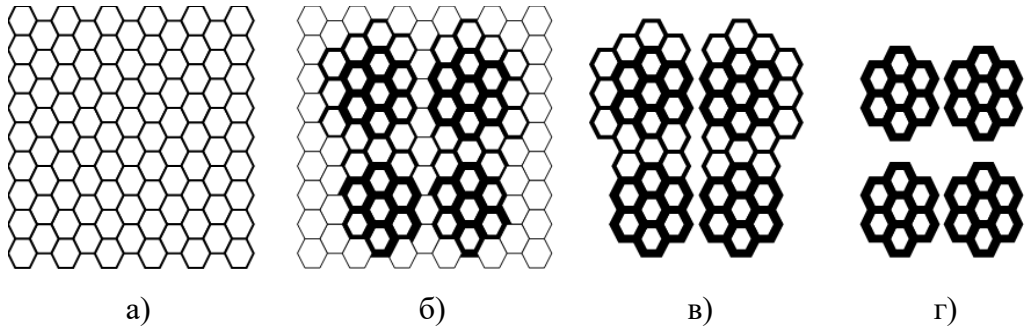

а) б) в) г)

Рис. 2. Застосування $\lambda$-серцевин для виділення спільнот у мережі

У даному пункті також визначаються умови найбільш уразливих до критичної завантаженості функціонально важливих складових МС. Елементи матриці $\mathbf{V}^\lambda(t)$ зі зростанням значення $\lambda$ визначають функціональну пріоритетність відповідних підсистем мережевих систем. Вводиться поняття $\beta$-серцевини завантаженості МС, як найбільшої підмережі вихідної мережі, для якої елементи матриці $\mathbf{U}_L(t)$ є не меншими за значення $\beta \in [0,1]$. Матриця суміжності $\beta$-серцевини $\mathbf{U}_L^\beta(t) = \{U_{L,ij}^\beta(t)\}_{i,j=1}^N$ визначається за співвідношенням

$$U_{L,ij}^\beta(t) = \begin{cases} U_{L,ij}(t), \text{ якщо } U_{L,ij}(t) \geq \beta, \\ 0, \text{ якщо } U_{L,ij}(t) < \beta, \quad t \geq T, i,j = \overline{1,N}. \end{cases}$$

Тоді ненульові елементи матриці

$$\mathbf{W}^{\lambda \times \beta}(t) = \{V_{ij}^\lambda(t) \times U_{L,ij}^\beta(t)\}_{i,j=1}^N$$

при значеннях $\lambda$ та $\beta$, близьких до 1, визначають найбільш уразливі із функціонально найважливіших складових системи, які належать перетину $\lambda$- та $\beta$-серцевин МС. Наведені приклади свідчать, що незначне збільшення об'ємів потоків у таких складових може призвести до їх блокування в системі, заподіявши їй найбільшу шкоду.

У пункті 3.7 вивчається поведінка мережевих систем з безмасштабною структурою на усіх етапах її життєвого циклу. Визначаються особливості розвитку таких систем, зокрема поведінка параметрів росту та переважного приєднання, які доповнюють модель Барабаші-Альберт, та вводяться критерії, які дозволяють визначити етап життєвого циклу, на якому перебуває система.

У пункті 3.8 досліджується проблема ізольованих зон мережі та визначаються показники, які на основі параметрів впливу та посередництва ізольованої зони дозволяють кількісно обчислювати втрати, які очікують систему у разі появи таких зон. Формулюються задачі, пов'язані з необхідністю перерозподілу потоків альтернативними шляхами в обхід ізольованих зон мережі та пропонуються алгоритми їх розв'язання.

У четвертому розділі **«Потокові моделі міжсистемних взаємодій»** вивчаються функціональні та структурні особливості багатошарових мережевих систем (БШС) різних типів.

У пункті 4.1 формулюються основні функціональні та структурні ознаки класифікації багатошарових мережевих систем. Серед функціональних ознак виділяються мульти- та монопотоковість системи, одно- та різнотипність носіїв потоків, рівень взаємодії шарів-систем та мультимережевість міжсистемних взаємодій. Серед основних структурних ознак класифікації БШС виділяються перетин сукупностей вузлів шарів-мереж та способи організації міжшарових взаємодій. Залежно від визначеного за виділеними вище ознаками типу БШС пропонуються способи побудови матриці суміжності її структури, які спрощують аналіз міжсистемних взаємодій різних видів.

У пункті 4.2 визначаються та досліджуються структурні характеристики монопотокових частково покритих БШС. Для систем цього типу вводиться поняття її агрегат-мережі, яке суттєво спрощує дослідження багатьох властивостей вихідної багатошарової мережі, та агрегат-ваги її вузлів та ребер. Для вузлів частково покритої БШМ визначаються їх вхідні та вихідні ступені та сили взаємозв'язку, а також ступені вхідних та вихідних міжшарових взаємозв'язків точок переходу.

У пункті 4.3 визначаються такі структурні характеристики шарів частково покритих багатошарових мереж, як питомі ваги сукупності вузлів та зв'язків окремого шару у загальній сукупності вузлів та зв'язків частково покритої БШС, вхідний та вихідний ступінь шару, який дорівнює кількості точок переходу на даний шар та з даного шару на всі інші шари БШС, а також питома вага точок переходу шару у сукупності усіх точок переходу частково покритої БШС, яка визначає доступність міжшарових взаємодій для даного шару.

Для вирішення проблеми визначення структурно найважливіших складових міжсистемних взаємодій у БШС введено поняття структурної $p$-серцевини частково покритої багатошарової мережі, як поєднання тих складових окремих шарів, які входять до складу не менше, ніж $p$, $2 \leq p \leq M$, шарів. Окремо виділено поняття $M$-серцевини або ядра БШС, яке дозволяє поділяти частково покриті багатошарові системи на ядерні (рис. 3) та без'ядерні.

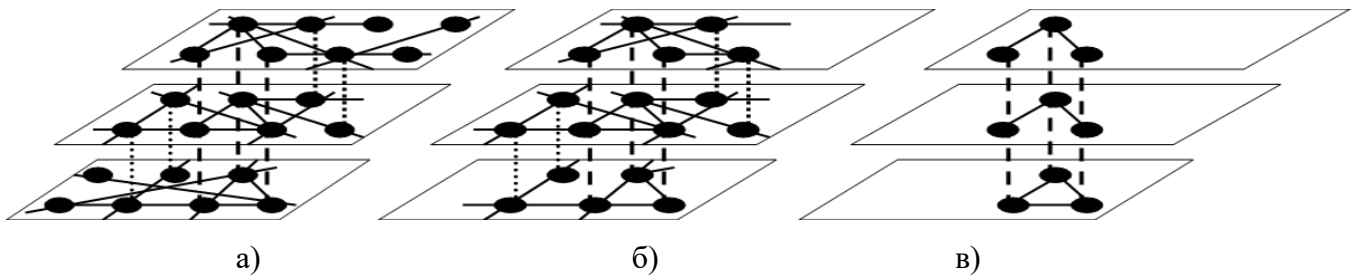

Рис. 3. Фрагмент ядерної БШС (а) та її *2*- (б) і *3*-серцевин (в)

Показується, що $p$-серцевини та агрегат-мережі БШС (рис. 4) суттєво спрощують дослідження міжсистемних взаємодій та визначають критерій для виділення спільнот у монопотокових БШС. Якщо $M$-серцевина БШС співпадає з її структурою, то відповідна багатошарова мережа є мультиплексом. Показується, що ефект «тісного світу» у мультиплекс-структурі, кожний шар якої володіє цією властивістю, посилюється у сенсі зменшення середньої довжини найкоротшого шляху та збільшення коефіцієнта кластеризації. У випадку послідовності зв'язних $p$-серцевин монопотокової ядерної або без'ядерної частково покритої БШС, кожний шар якої є мережею «тісного світу», цей ефект також буде послідовно посилюватися зі зростанням значення $p$ у кожній із $p$-серцевин, $2 \leq p \leq M$.

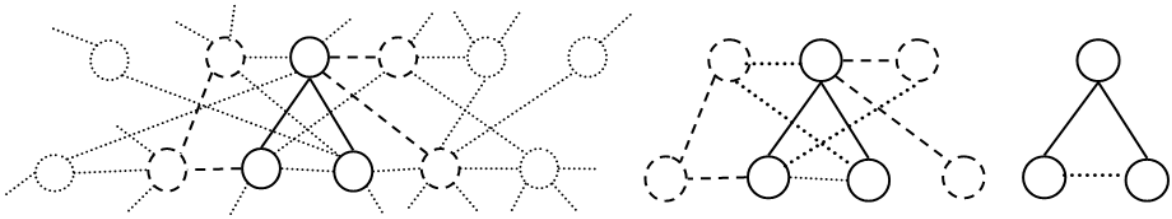

Рис. 4. Фрагмент зваженої агрегат-мережі, зображеної на рис. 3 ядерної БШМ (а) та її *2*- (б) і *3*-серцевин (в) (——— – елемент входить до складу трьох шарів, - - - - – елемент входить до складу двох шарів, ……. – елемент входить до складу одного шару)

У пункті 4.3 побудована потокова модель монопотокових частково покритих багатошарових систем

$$\mathbf{P}^M(t,x) = \{\boldsymbol{\rho}^M(t,x), \mathbf{v}^M(t), \mathbf{u}^M(t), t \geq 0; \mathbf{V}^M(t), \mathbf{U}^{M,C}(t), \mathbf{U}^{M,L}(t), t \geq T\},$$

у якій $\boldsymbol{\rho}^M(t,x)$ – матриця щільності потоків, які пересуваються ребрами багатошарової мережі в поточний момент часу $t$; $\mathbf{v}^M(t)$ – поточна матриця суміжності об'ємів потоків, які знаходяться на ребрах БШМ в момент часу $t$; $\mathbf{u}^M(t)$ – матриця завантаженості ребер багатошарової мережі в момент часу $t$, $t \geq 0$; $\mathbf{V}^M(t)$ – інтегральна потокова матриця суміжності об'ємів потоків, які пройшли ребрами БШС за період $[t-T,t]$ до поточного моменту часу $t$; $\mathbf{U}^{M,C}(t), \mathbf{U}^{M,L}(t)$ – інтегральні матриці завантаженості багатошарової мережі за період $[t-T,t]$ до моменту часу $t$, $t \geq T$.

Визначаються потокові агрегат-ваги та агрегат-ступені вузлів БШС, які встановлюють локальну важливість елементів БШС. Для визначення глобальної важливості елементів у процесі міжсистемних взаємодій вводяться параметри впливу та посередництва вузлів та ребер БШС, які дозволяють визначати важливість шляхів, якими реалізуються міжсистемні взаємодії, значення вузлів – генераторів, приймачів та транзитерів міжсистемних потоків, параметри переважного впливу вузлів БШС у процесі міжсистемних взаємодій, а також яким чином блокування точок переходу впливатиме на процес функціонування областей їх посередництва, величину цих областей і, унаслідок цього, – окремого шару-системи або БШС загалом. Міра посередництва та процес перерозподілу руху потоків дозволяє визначити втрати, які у результаті очікують окремий шар або багатошарову систему в цілому.

У пункті 4.4 визначаються та досліджуються найважливіші потокові характеристики шарів-систем у монопотоковій частково покритій багатошаровій мережевій системі, до яких відносяться: обсяги внутрішньошарових потоків в окремому шарі БШС; обсяги вихідних та вхідних потоків шару, які відображають його міжсистемні взаємодії у багатошаровій системі; співвідношення між обсягами внутрішньо та міжсистемних потоків шару; інтегральні обсяги внутрішньошарових та міжшарових взаємодій у БШС; співвідношення між внутрішньо та міжсистемними взаємодіями у БШС загалом; інтегральний показник процесу функціонування БШС, який визначає сумарні обсяги потоків у монопотоковій частково покритій багатошаровій системі протягом періоду часу $[t-T,t], t \geq T$. Також вводяться параметри впливу та посередництва шарів монопотокових частково покритих багатошарових систем, за допомогою яких визначається важливість окремих шарів БШС у процесі міжсистемних взаємодій, їх роль, як генераторів, приймачів та транзитерів міжсистемних потоків і т. ін. Ці параметри є глобальними характеристиками важливості шару у процесі міжсистемних взаємодій та визначають, яким чином блокування шару вплине на процес функціонування області його посередництва, величину цієї області і, унаслідок цього, – інших шарів-систем або всієї монопотокової частково покритої БШС.

Пропонуються два способи побудови потокових $\lambda$-серцевин монопотокових частково покритих БШС. Перший із них полягає у виділенні в окрему багатошарову структуру тих ребер та точок переходу БШС, об'єми руху потоків через які за проміжок часу $[t-T,t], t \geq T$, є не меншими значення $\lambda$, $\lambda \in [0,1]$. Другий спосіб зводиться до визначення тих ребер та точок переходу, сумарні об'єми руху потоків через які в усіх шарах БШС, до складу яких вони входять, за той же проміжок часу є не меншими, ніж $\lambda$. На рис. 5 відображені фрагмент структури ядерної монопотокової частково покритої БШС (а), її $3$-серцевини або ядра (б), **4**={4, 4, 4}-серцевини (в) та потокової $\lambda$-серцевини, побудованої другим із запропонованих вище способів (г). Очевидно, що $p$-серцевини дають значно важливішу інформацію для аналізу особливостей структури, а $\lambda$-серцевини – для дослідження процесу функціонування багатошарових мережевих систем, ніж їх **k**-серцевини.

У пункті 4.5 пропонуються методи ідентифікації фіктивних та пошуку прихованих елементів БШС, що дозволяє визначати реальну структуру внутрішньо та міжсистемних взаємодій, алгоритми виділення спільнот у БШС за допомогою потокових серцевин другого типу (при цьому найбільш цікавим є випадок, коли спільноти є багатошаровими утвореннями, внутрішньо та міжшарові зв'язки між елементами яких є щільнішими та з функціонального погляду «сильнішими», ніж у середньому для БШС). Досліджується проблема стійкості багатошарових мережевих систем, яку також поділено на визначення їх уразливості до цілеспрямованих атак та умов критичної завантаженості окремих складових БШС. Уразливість до негативних впливів на процес міжсистемних взаємодій у БШС визначається за допомогою параметрів посередництва точок переходу багатошарової системи. Уразливість міжшарових зв'язків до умов критичної завантаженості безпосередньо залежить від пропускної здатності існуючих точок переходу БШС.

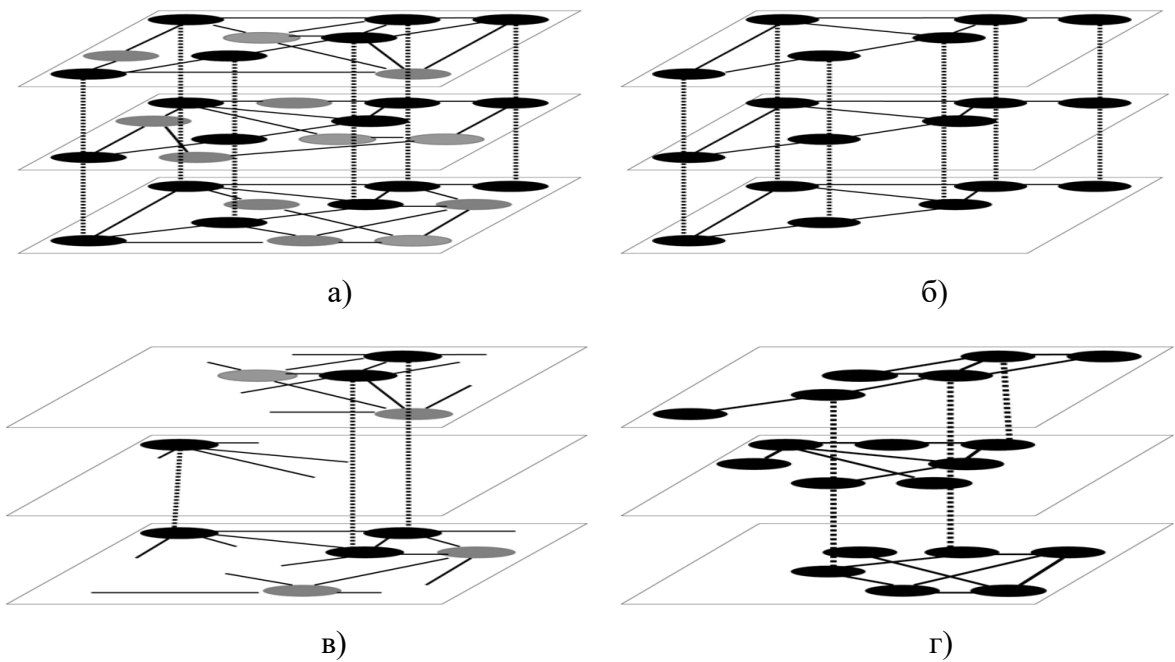

Рис. 5. Фрагмент структури ядерної тришарової частково покритої БШС (а), її ядра (б), структурної (в) та потокової (г) серцевин

У п'ятому розділі **«Складні ієрархічно-мережеві системи»** досліджуються структурні та функціональні особливості складних систем з ієрархічно-мережевою структурою, сформованою за принципом прямого підпорядкування та лінійною моделлю управління.

У пункті 5.1 кожна СІМС поділяється на дві основні складові:
1) шар-систему найнижчого рівня ієрархії, яка зазвичай реалізує ціль утворення та існування СІМС, тобто забезпечує рух певного типу потоків – транспортних, ресурсних, фінансових, інформаційних тощо;
2) ієрархічно-мережеву систему управління (СУ), яка за допомогою управлінських, організаційних, інформаційних та інших потоків повинна забезпечувати ефективне функціонування системи найнижчого рівня ієрархії, зокрема, оперативне реагування як на негативні, так і на позитивні зміни, аналіз та прогнозування цих змін, своєчасне попередження розвитку загрозливих тенденцій і т. ін.

Оскільки множини вузлів шарів-мереж у СІМС практично не перетинаються, то формально вони мають мультимережеву структуру. Якщо вважати, що основним за своєю важливістю видом потоків у СУ є управлінські рішення та реакції на них, то СІМС прямого підпорядкування є мультипотоковою БШС, два основні види потоків якої зосереджені на нижньому та сукупності вищих рівнів ієрархії системи відповідно. Загалом структура СІМС утворює свого роду «піраміду», в основі якої лежить шар, заради якого система створювалась та функціонує. Тому цей шар найнижчого рівня ієрархії називатимемо системою-основою (рис. 6).

У пункті 5.2 сформована структурна модель СІМС прямого підпорядкування у вигляді відповідної матриці суміжності, а у пункті 5.3 визначені основні локальні та глобальні внутрішньо та міжшарові характеристики складових ієрархічно-мережевих структур.

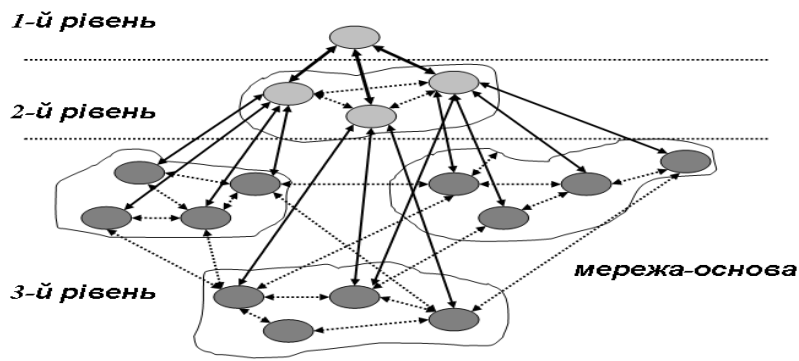

Рис. 6. Приклад системи з трирівневою ієрархічно-мережевою структурою прямого підпорядкування

У пункті 5.4 побудована потокова модель складної ієрархічно-мережевої системи та визначено внутрішньо та міжшарові ступені вузлів СІМС, а також параметри впливу та посередництва елементів системи. Визначаються основні характеристики підсистем-шарів та ієрархічно-мережевих підсистем СІМС, які встановлюють їх важливість у процесі функціонування системи. Вводиться поняття потокової $\lambda$-серцевини СІМС та досліджується її застосування для зменшення розмірності моделі системи та визначення шляхів оптимізації структури СУ. Перший із цих шляхів полягає у видаленні мало завантажених елементів управлінських рівнів ієрархії та перерозподілу їх функцій між суміжними із видаленими керуючими вузлами до досягнення ними порогу критичної завантаженості. Другий шлях полягає в інкапсуляції мало завантажених зв'язних складових управлінських шарів СІМС та їх заміщенні мета-вузлами, завантаженість яких буде співмірною з іншими керуючими вузлами системи.

У пункті 5.5 досліджується проблема стійкості складних ієрархічно-мережевих систем, яка поділяється на дві взаємопов'язані проблеми – уразливості системи-основи та системи управління СІМС. Проблема стійкості системи-основи поділяється на проблему уразливості до цілеспрямованих атак та проблему критичної завантаженості або чутливості до малих змін в структурі або процесі функціонування мережевої системи. Основне завдання вузлів СУ полягає у забезпеченні дієвого захисту елементів підпорядкованих їм підсистем системи-основи, а у разі неспрацювання цього захисту – в оперативному відновленні процесу функціонування уражених елементів та мінімізації наслідків цього ураження. При цьому, чим більша зона ураження системи-основи, тим вищий рівень управління повинен бути задіяний для подолання наслідків цих уражень та протидії повторним атакам. У функцію керуючих вузлів СІМС також входить упередження умов критичної завантаженості підпорядкованих їм підсистем системи-основи, а у разі появи таких умов – максимально швидке розвантаження цих підсистем.

Проблема стійкості елементів системи управління СІМС до цілеспрямованих атак поділяється на три взаємопов'язані складові: атаки на вузли певного управлінського шару, спрямовані на блокування внутрішньо шарових керуючих взаємодій; атаки на вузол певного управлінського шару з метою блокування підпорядкованої йому ієрархічно-мережевої підсистеми СІМС; атаки на міжшарові зв'язки системи управління СІМС. Запропоновані принципи побудови сценаріїв вищеперерахованих видів атак є найдієвішими з погляду ураження найважливіших для процесу функціонування елементів як системи-основи, так і системи управління СІМС. При цьому чим вищий рівень ураження, тим потенційно більші втрати очікують систему. Кількісно ці втрати визначаються важливістю вузла системи управління, способи обчислення якої визначаються локальними та глобальними потоковими характеристиками вузла в системі. Досліджується випадок, коли елементи-цілі атак на систему є різними (елементи системи-основи або системи управління), але наслідки цих атак, зокрема зони ураження, можуть бути подібними. Визначаються також внутрішні причини, які можуть уразити складові системи управління СІМС, серед яких особливо виділяються умови критичного завантаження її вузлів інформаційними потоками, які фізично неможливо опрацювати у визначені проміжки часу та можуть викликати ефект «паралічу аналізу», а також несинхронізоване надходження потоків даних у певний вузол управління, який унеможливлює своєчасне прийняття правильного рішення унаслідок недостатності необхідної інформації.

У пункті 5.6 описуються принципи формування інформаційної моделі СІМС, на основі якої сумісно з динамічними структурами пріоритетності та наповненості даними будуються моделі регулярного та інтерактивного оцінювання стану та якості функціонування складних ієрархічно-мережевих систем. Основна ціль створення інформаційної моделі – спрямованість на відповідні рівні та елементи управління СІМС для спрощення процесу аналізу даних та прийняття рішень щодо подальших дій стосовно підпорядкованих їм складових системи. Модель регулярного оцінювання будується на основі інформації, отриманої під час періодичних планових досліджень СІМС або зібраної протягом певного часу її функціонування, та передбачає:
- локальне оцінювання стану, якості функціонування та взаємодії елементів (вузлів та ребер) системи-основи СІМС;
- агреговане оцінювання елементів системи на усіх управлінських рівнях ієрархії, при якому узагальнений висновок про стан та якість функціонування підпорядкованої певному елементу підмережі СІМС є визначальним для оцінки цього елемента системи;
- прогностичне оцінювання стану та якості функціонування елементів усіх рівнів ієрархії СІМС.

Метою регулярного оцінювання є глибокий та ретельний аналіз стану, процесу функціонування та взаємодії усіх елементів системи. Структура агрегованих оцінок є тотожною до структури оцінюваної СІМС, а термін прогнозу не може бути меншим за необхідний для усунення виявлених загроз.

Модель інтерактивного оцінювання СІМС будується на основі результатів неперервного моніторингу процесу функціонування системи, які містяться в інформаційній моделі, та передбачає:

- локальне оцінювання взаємодії потоків та елементів структури (вузлів та ребер) у системі-основі СІМС;
- агреговане оцінювання взаємодії елементів системи на усіх рівнях ієрархії, при якому узагальнений висновок про якість взаємодії елементів підпорядкованої певному елементу підмережі СІМС є визначальним для оцінки цього елемента;
- прогностичне оцінювання якості взаємодії елементів усіх рівнів ієрархії СІМС.

Інтерактивне оцінювання проводиться неперервно у режимі реального часу та полягає у постійному відстежуванні взаємодії мережевих та міжрівневих потоків з вузлами та ребрами СІМС. Висновки, отримані у результаті інтерактивного оцінювання, є опосередкованими, але від того не менш важливими для контролю за станом та якістю функціонування системи і окремих її складових. Структури наповненості та пріоритетності у моделях оцінювання визначають рівень покриття оцінками елементів СІМС усіх рівнів ієрархії залежно від наявних в інформаційній моделі даних та пріоритетність аналізу цих оцінок відповідно.

У шостому розділі **«Локальне оцінювання поведінки елементів складних систем»** досліджуються особливості багатокритеріального та багатопараметричного оцінювання стану та якості функціонування елементів складних систем.

У пункті 6.1 на реальних прикладах розглядаються основні особливості оцінювання елементів системи, яке починається з формування набору характеристик, що описують їх найважливіші властивості, та передбачає насамперед аналіз стану елемента, а потім якості реалізації покладених на нього функцій, які прямо чи опосередковано від цього стану залежать.

У пункті 6.2 здійснюється порівняльний аналіз неперервної, цілочисельної та понятійної шкал оцінювання характеристик елементів системи та пропонується багатопараметрична уточнена бальна шкала оцінок, а саме, уточнена бальна оцінка $E_V(f)$ характеристики $f(x)$, $x \in [0, X]$, де $x$ – просторова або часова змінна, визначається за співвідношенням:

$$E_V(f) = \begin{cases} e(f), & \text{якщо } i = 2, \\ e(f) + (1 - \|P_{F_i}(f)\|_V / \upsilon_i), & \text{якщо } i = 3,4, \\ e(f), & \text{якщо } i = 5. \end{cases}$$

Тут $i = 2, 3, 4, 5$ – основна бальна оцінка характеристики $f(x)$, $P_{F_i}(f)$ – проекція на підобласть $F_i[0, X]$ значень характеристики $f(x)$, основна бальна оцінка якої дорівнює $i$, $\upsilon_i$ – нормуючий коефіцієнт, $i = 3, 4$, $\|.\|_V$ – норма функціонального простору $V$. Показано, що уточнена бальна шкала оцінювання поєднує переваги неперервної та понятійної шкал, а саме точність і зрозумілість кінцевому користувачу та дозволяє принаймні частково локалізувати причини виявлених недоліків. У пункті описуються алгоритми уточненого бального оцінювання дискретних характеристик елемента системи, а також поведінки його параметрів впливу та посередництва.

У пункті 6.3 пропонуються методи інтерактивного оцінювання результатів неперервного моніторингу елементів системи, які застосовуються для МС із повністю впорядкованим рухом потоків. Описується алгоритм інтерактивного оцінювання, який ґрунтується на аналізі взаємодії таких елементів системи, як потік та ребро або вузол мережі. Пропонуються методи інтерактивного оцінювання негативних та позитивних впливів, які 21можуть здійснювати вузли МС на систему і навпаки, а також визначаються елементи, які безумовно підлягають неперервному моніторингу у довільних реальних складних системах.

У пункті 6.4 аналізується ефективність існуючих та пропонуються нові методи побудови узагальнених висновків про якість елемента системи. Нехай стан та/або процес функціонування елемента описується сукупністю характеристик $\mathbf{\Phi} = \{f_n\}_{n=1}^{N}$ та $e(f_n)$ – оцінка поведінки характеристики $f_n$, $n = \overline{1,N}$, отримана з використанням уточненої бальної шкали оцінок. Якість елемента системи $E_{me}$ за першим методом визначається найменшою оцінкою характеристики із сукупності $\mathbf{\Phi}$, тобто

$$E_{me} = \min_{n=1,N}\{e(f_n)\}.$$

Цей метод, який називатимемо методом найменшої оцінки (МНО), використовується під час оцінювання елементів, у яких незадовільна поведінка однієї з характеристик може призвести до припинення їх функціонування.

Метод зваженої лінійної агрегації (МЗЛА) визначається за співвідношенням

$$E_{wla}(\mathbf{\Phi},\mathbf{\rho}) = <\mathbf{\rho}, \mathbf{e}(\mathbf{\Phi})>_{R^N} / <\mathbf{\rho}, \mathbf{1}>_{R^N},$$

у якому $\mathbf{e}(\mathbf{\Phi}) = \{e(f_n)\}_{n=1}^{N}$, $\mathbf{\rho} = \{\rho_n\}_{n=1}^{N}$ – вектор вагових коефіцієнтів, який встановлює пріоритетність характеристик елемента системи і $\mathbf{1} = \{1\}_{n=1}^{N}$. Основним недоліком МЗЛА є нівелювання як позитивних, так і негативних оцінок характеристик елемента.

Нехай ми маємо сукупність взаємопов'язаних однотипних елементів системи $\mathbf{Q}_N = \{q_n\}_{n=1}^{N}$ стан або процес функціонування кожного з яких описується однаковим набором характеристик $\mathbf{F}_K^n = \{f_k^n\}_{k=1}^{K}$, $n = \overline{1,N}$. Побудова узагальненого виновку про поведінку складової $\mathbf{Q}_N$ системи здійснюється у двох напрямках, які умовно називаємо вертикальним та горизонтальним. Побудова узагальненого виновку у горизонтальному напрямку передбачає:

1) зважену лінійну агрегацію оцінок $\mathbf{e}(\mathbf{F}_K^n) = \{e(f_k^n)\}_{k=1}^{K}$ характеристик елемента $q_n$, $n = \overline{1,N}$, тобто

$$H_n(\mathbf{F}_K^n) = <\mathbf{\rho}_K, \mathbf{e}(\mathbf{F}_K^n)>_{R^K} / <\mathbf{\rho}_K, \mathbf{1}>_{R^K},$$

де $\mathbf{\rho}_K = \{\rho_k\}_{k=1}^{K}$ – вектор вагових коефіцієнтів, які визначають пріоритетність характеристик елемента;

2) зважену лінійну агрегацію оцінок елементів, які входять до складу $\mathbf{Q}_N$, за сукупністю характеристик, тобто

$$H(\mathbf{Q}_N) = <\mathbf{r}_N, \mathbf{H}(\mathbf{F}_K^N)>_{R^N} / <\mathbf{r}_N, \mathbf{1}>_{R^N},$$

де $\mathbf{H}(\mathbf{F}_K^N) = \{H_n(\mathbf{F}_K^n)\}_{n=1}^{N}$ та $\mathbf{r}_N = \{r_n\}_{n=1}^{N}$ – вектор вагових коефіцієнтів, які визначають пріоритетність елементів у сукупності $\mathbf{Q}_N$.

Побудова узагальненого висновку у вертикальному напрямку передбачає:

1) зважену лінійну агрегацію оцінок характеристик $f_k^n$, $k = \overline{1,K}$, для усіх елементів, які входять до складу $\mathbf{Q}_N$, за сукупністю елементів, тобто

2222$$V_k(\mathbf{F}_k^N) = <\mathbf{r}_N, \mathbf{e}(\mathbf{F}_k^N)>_{R^N} / <\mathbf{r}_N, \mathbf{1}>_{R^N},$$

де $\mathbf{e}(\mathbf{F}_k^N) = \{e(f_k^n)\}_{n=1}^N$, $\mathbf{F}_k^N = \{f_k^n\}_{n=1}^N$;

2) зважену лінійну агрегацію оцінок характеристик елементів сукупності $\mathbf{Q}_N$, тобто

$$V(\mathbf{Q}_N) = <\boldsymbol{\rho}_K, \mathbf{V}(\mathbf{F}_K^N)>_{R^K} / <\boldsymbol{\rho}_K, \mathbf{1}>_{R^K},$$

де $\mathbf{V}(\mathbf{F}_K^N) = \{V_k(\mathbf{F}_k^N)\}_{k=1}^K$.

Оцінювання у вертикальному напрямку насамперед формує узагальнений висновок про якість кожного елемента за сукупністю його характеристик, а потім – про якість елементів сукупності $\mathbf{Q}_N$ загалом. Таким чином воно дозволяє визначити проблемні елементи, що входять до складу $\mathbf{Q}_N$. Оцінювання у горизонтальному напрямку спочатку формує узагальнений висновок про поведінку окремої характеристики для всіх елементів сукупності $\mathbf{Q}_N$, а потім – про якість елементів сукупності в цілому. Таким чином воно дозволяє визначити ті характеристики елементів, що входять до складу $\mathbf{Q}_N$, поведінка яких є найгіршою. Доведена

**Теорема 6.1.** Узагальнений висновок $E(\mathbf{Q}_N)$ про стан або якість функціонування складової системи $\mathbf{Q}_N$, отриманий методом зваженої лінійної агрегації, визначається співвідношенням

$$E(\mathbf{Q}_N) \equiv H(\mathbf{Q}_N) = V(\mathbf{Q}_N).$$

У випадку рівної важливості характеристик елемента більш адекватний до дійсності результат узагальнення порівняно із МЗЛА отримуємо за допомогою методу нелінійної агрегації (МНА), який визначається за співвідношенням

$$E_{na}(\boldsymbol{\Phi}) = \prod_{n=1}^N e(f_n)/(e^*)^{N-1}, \quad e^* = \sum_{n=1}^N e(f_n)/N.$$

Доведена

**Теорема 6.2.** Для довільного набору оцінок $\{e(f_n)\}_{n=1}^N$ сукупності рівно важливих характеристик $\boldsymbol{\Phi}$ елемента системи виконуються нерівності

$$E_{me}(\boldsymbol{\Phi}) \leq E_{na}(\boldsymbol{\Phi}) \leq E_{la}(\boldsymbol{\Phi}).$$

При цьому, якщо $e(f_n) = E_{me}(\boldsymbol{\Phi})$, то має місце рівність

$$E_{me}(\boldsymbol{\Phi}) = E_{na}(\boldsymbol{\Phi})$$

та, якщо $e(f_n) = e^*$, $n = \overline{1,N}$, – рівність

$$E_{na}(\boldsymbol{\Phi}) = E_{la}(\boldsymbol{\Phi}).$$

Для покращення якості функціонування елементів системи пропонуються ефективні способи послідовного застосування методів найменшої оцінки та зваженої лінійної або нелінійної агрегації. Для поєднання переваг МЗЛА і МНА розроблено метод гібридної агрегації (МГА), який визначається співвідношенням

$$E_{ha}(\boldsymbol{\Phi}, \boldsymbol{\rho}) = <\boldsymbol{\rho}, E_{na}(\boldsymbol{\Phi})>_{R^M} / <\boldsymbol{\rho}, \mathbf{1}>_{R^M},$$

де $E_{na}(\mathbf{\Phi}) = \{E_{na}(\mathbf{\Phi}_m)\}_{m=1}^{M}$, $\mathbf{\Phi}_m = \{f_n\}_{n=1}^{n_m}$, $n_m \geq 1$, – групи характеристик, які мають однакову пріоритетність $\rho_m$, $m = \overline{1, M}$, та $\mathbf{\rho} = \{\rho_m\}_{m=1}^{M}$. Доведена

**Теорема 6.3.** Виконується наступна нерівність

$$E_{me}(\mathbf{\Phi}) \leq E_{ha}(\mathbf{\Phi}, \mathbf{\rho}) \leq E_{wla}(\mathbf{\Phi}, \mathbf{\rho}^*),$$

де $\mathbf{\rho}^* = \{\{\rho_m\}_{n=1}^{n_m}\}_{m=1}^{M}$.

На реальних прикладах показується, що МГА дає значно адекватніші узагальнені висновки порівняно з іншими методами агрегації.

У пункті 6.5 пропонуються методи прогнозування поведінки оцінок елементів системи. Показано, що розроблені на основі екстраполяційних підходів алгоритми прогнозування уточнених бальних оцінок дають можливість визначити момент часу, коли понятійна оцінка характеристики елемента зменшиться на одиницю, наприклад, досягне значення «незадовільно», завчасно скорегувати (наблизити або віддалити) плановий термін наступного огляду системи, здійснювати середньо- та довгострокове прогнозування поведінки оцінок характеристик елемента, уникаючи більш складних для чисельної реалізації методів часових рядів тощо. Результати інтерактивного оцінювання використовуються для більш детального і точного прогностичного аналізу роботи оцінюваного елемента системи та дозволяють своєчасно корегувати прогнози, здійснені на основі його регулярних досліджень.

У сьомому розділі **«Комплексне оцінювання поведінки складних мережевих та ієрархічно-мережевих систем»** методи агрегованого оцінювання застосовуються для побудови узагальнених висновків стосовно якості функціонування окремих підсистем МС, вибору оптимальних режимів функціонування системи та оптимальної системи з даного класу еквівалентних систем, аналізу історії функціонування МС та міжсистемних взаємодій різних типів.

У пункті 7.1 оцінка окремої підсистеми МС будується на основі підбору найкращої гібридної агрегаційної схеми, яка б давала найбільш адекватний узагальнений висновок стосовно ефективності її роботи. У якості прикладу використовується аналіз процесу функціонування ліній руху потоків мережею. Також оцінюються потенційні позитивні та негативні впливи окремої підсистеми на МС загалом і навпаки.

У пункті 7.2 на основі методу гібридної агрегації пропонуються алгоритми вибору оптимальних та визначення екстремальних режимів функціонування системи, а у пункті 7.3 – вибору оптимальної системи із даного класу еквівалентних систем. Аналізуються процеси, які розгортаються в МС протягом періоду її існування з метою ідентифікації етапу життєвого циклу, на якому перебуває система, та розробляються рекомендації щодо подальших дій стосовно неї.

У пункті 7.4 формуються узагальнені висновки стосовно міжсистемних взаємодій, які виникають у моно- та мультипотокових багатошарових системах різних типів. Досліджуються процеси, які можуть розвиватися у таких надсистемних утвореннях, як асоціації, конгломерати та системні середовища.

У пункті 7.5 досліджується проблема пошуку «аномалій» в інформаційних моделях складних систем, актуальність вирішення якої у сучасному світі постійно зростає із неперервним збільшенням обсягів даних, що накопичуються у багатьох реальних СІМС. На основі застосування методів теорії оцінювання пропонуються підходи, які дозволяють суттєво спростити вирішення цієї проблеми принаймні в моделях структурованих даних про систему.

У пункті 7.6 на прикладі автотранспортної системи (АТС) великого міста пропонуються методи оцінювання процесу функціонування мережевих систем із частково впорядкованим рухом потоків. У якості впорядкованої частини потоків обирається обладнаний GPS-трекерами громадський транспорт, рух якого підпорядкований певному графіку, дотримання якого можна неперервно моніторити. Для інтерактивного оцінювання використовується метод U-статистик, на основі якого формуються уточнені бальні оцінки стану та ефективності функціонування елементів АТС, які дозволяють аналізувати та прогнозувати автотранспортні ситуації на окремих ділянках міської автомережі. Для формування узагальнених висновків про стан, ефективність функціонування та розвиток автотранспортних ситуацій на окремих маршрутах руху громадського транспорту, у регіонах міста та його АТС загалом застосовуються методи нелінійної та гібридної агрегації. Як і для систем із повністю впорядкованим рухом потоків, запропонована у даному пункті методика також має комплексний характер, оскільки поєднує у собі взаємопов'язані методи інтерактивного, прогностичного та агрегованого оцінювання поведінки складових системи різного типу та рівнів ієрархії.

У **висновках** наведено основні результати дисертаційної роботи.

У **додатках** наведено опис основних типів поведінки неперервних характеристик поведінки елементів складних систем (Додаток А), принципи візуалізації результатів комплексного оцінювання системи різних рівнів узагальнення (Додаток Б), перелік опублікованих за темою дисертації наукових праць (Додаток В) та подано акти про використання результатів дисертації (Додаток Г).

## ВИСНОВКИ

На основі виконаних теоретичних та експериментальних досліджень у дисертаційній роботі вирішено важливу науково-прикладну проблему визначення функціональних властивостей складних мережевих та багатошарових мережевих систем різних типів та розроблення методики комплексного оцінювання стану, процесу функціонування та взаємодії елементів таких систем, як математичного підґрунтя для створення засобів підтримки прийняття рішень різного призначення.

Основні наукові та практичні результати, отримані в роботі, полягають в наступному.

1. Проведено аналіз сучасного стану дослідження складних мережевих систем та міжсистемних взаємодій різних типів, виділено основні напрямки цих досліджень та показано, що вони загалом обмежуються вивченням структурних особливостей таких систем та взаємодій. Розглянуто основні підходи до оцінювання складних систем та показано доцільність розробки комплексної методики оцінювання, яка поєднує методи локального, інтерактивного, прогностичного та агрегованого аналізу стану та якості функціонування складових системи.

2. Розроблено потокові моделі складних мережевих систем і визначено локальні та глобальні потокові характеристики їх елементів, які застосовано для встановлення реальної структури системи, аналізу її розвитку на усіх етапах життєвого циклу та пошуку альтернативних шляхів руху потоків в обхід ізольованих зон мережі.

3. Визначено основні структурні та функціональні ознаки класифікації міжсистемних взаємодій різних типів та показано їх вплив на процес функціонування пов'язаних між собою складних мережевих систем.

4. Визначено параметри впливу та посередництва складових мережевих та монопотокових частково покритих багатошарових мережевих систем, з використанням яких досліджені умови стійкості системи до негативних внутрішніх та зовнішніх впливів різних типів, встановлено важливість різнорідних внутрішньо та міжсистемних взаємодій і запропоновано сценарії цілеспрямованих атак на мережеві системи з метою визначення найбільш привабливих цілей таких атак та розроблення відповідних засобів їх захисту.

5. Запропоновано методи редукції складності мережевих систем та міжсистемних взаємодій різних типів, які дозволяють будувати адекватні моделі таких систем та взаємодій, придатні для практичного розв'язання задач керованості, спостережуваності та синхронізації і вирішення низки інших прикладних проблем системного аналізу та теорії прийняття оптимальних рішень.

6. Визначено поняття потокової серцевини складних мережевих та монопотокових частково покритих багатошарових мережевих систем, яка дозволяє виділяти в структурі системи найважливіші з функціонального погляду її складові, суттєво зменшувати розмірність моделі системи з одночасним відстеженням міри збереження її адекватності та вирішувати проблему пошуку спільнот у мережі.

7. Визначено поняття $p$-серцевини та ядра монопотокової частково покритої багатошарової мережі, які дозволяють виділяти найважливіші для реалізації міжсистемних взаємодій складові її структури та суттєво зменшувати розмірність відповідних моделей.

8. Визначено основні функціональні властивості складних систем з ієрархічно-мережевою структурою та розроблено методику багатокритеріального та багатопараметричного комплексного оцінювання стану та якості функціонування таких систем, яка поєднує взаємопов'язані методи локального, прогностичного, інтерактивного та агрегованого аналізу поведінки складових системи усіх рівнів ієрархії.

9. Розроблено уточнену бальну шкалу оцінювання поведінки характеристик елементів системи різних типів, яка дозволяє формувати значно точніші висновки про стан та процес функціонування елемента та принаймні частково локалізувати причини виявлених недоліків.

10. Розроблено методи інтерактивного оцінювання поведінки елементів системи, які на основі результатів неперервного моніторингу руху потоків мережею дозволяють у режимі реального часу формувати опосередковані, але від того не менш обґрунтовані висновки про стан складових системи та якість їх функціонування.

11. Розроблено метод нелінійного агрегованого оцінювання та ефективні способи гібридизації відомих агрегаційних процедур, які залежно від типу досліджуваної системи, дозволяють отримувати значно адекватніші узагальнені висновки про стан та процес функціонування її складових різних рівнів ієрархії.

12. Показано ефективність застосування розробленої методики комплексного оцінювання для вирішення проблеми визначення практично досяжних критеріїв якості елементів мережевих систем, пошуку новизни, визначення критичних і вибору оптимальних режимів функціонування системи та оптимальної системи з даного класу еквівалентних систем.

13. Розроблені методи комплексного оцінювання та принципи візуалізації отриманих висновків значно точніше за відомі локалізують складові системи, які потребують першочергового удосконалення стану або процесу функціонування, дозволяють своєчасно передбачати загрозливі тенденції їх розвитку і адекватно планувати витрати, необхідні для упередження існуючих або потенційних ризиків.

14. Розроблена в дисертаційній роботі методологія комплексного оцінювання може бути використана під час розробки експертних систем та систем підтримки прийняття рішень різного призначення у різних предметних областях та застосовувалась для оцінювання стану та якості функціонування біомеханічних і робототехнічних систем локомоційного типу, окремих підсистем залізничної транспортної системи різного рівня ієрархії, а також для аналізу ефективності організації навчального процесу у вищих навчальних закладах, що підтверджується відповідними актами.

# СПИСОК ОПУБЛІКОВАНИХ ПРАЦЬ ЗА ТЕМОЮ ДИСЕРТАЦІЇ

monitoring, decision support.